\documentclass[aps,prb,reprint,twocolumn,superscriptaddress,showpacs]{revtex4-1}
\usepackage{amsmath}
\usepackage{amssymb}
\usepackage{braket}
\usepackage{graphicx}
\usepackage{hyperref}
\usepackage{color}
\usepackage{multirow, makecell}
\usepackage{mathtools}
\usepackage{booktabs}

\AtBeginDocument{
	\heavyrulewidth=.08em
	\lightrulewidth=.05em
	\cmidrulewidth=.03em
	\belowrulesep=.65ex
	\belowbottomsep=0pt
	\aboverulesep=.4ex
	\abovetopsep=0pt
	\cmidrulesep=\doublerulesep
	\cmidrulekern=.5em
	\defaultaddspace=.5em
}

\newcommand{\abs}[1]{\left| #1 \right|}

\begin{document}

\title{Tunable effective length of fractional Josephson junctions}

\author{Daniel Frombach}
\affiliation{Institut f\"ur Mathematische Physik, Technische Universit\"at Braunschweig, D-38106 Braunschweig, Germany}

\author{Patrik Recher}
\affiliation{Institut f\"ur Mathematische Physik, Technische Universit\"at Braunschweig, D-38106 Braunschweig, Germany}
\affiliation{Laboratory for Emerging Nanometrology Braunschweig, D-38106 Braunschweig, Germany}

\begin{abstract}
Topological Josephson junctions (TJJs) have been a subject of widespread interest due to their hosting of Majorana zero modes. In long junctions, i.e. junctions where the junction length exceeds the superconducting coherence length, TJJs manifest themselves in specific features of the critical current \cite{Beenakker2013}. Here we propose to couple the helical edge states mediating the TJJ to additional channels or quantum dots, by which the effective junction length can be increased by tunable parameters associated with these couplings, so that such measurements become possible even in short junctions. Besides effective low-energy models that we treat analytically, we investigate realizations by a Kane-Mele model with edge passivation and treat them numerically via tight binding models. In each case, we explicitly calculate the critical current using the Andreev bound state spectrum and show that it differs in effective long junctions in the cases of strong and weak parity changing perturbations (quasiparticle poisoning).
\end{abstract}

\date{\today}
\maketitle

\section{Introduction}
\label{sec:introduction}

Topological Josephson junctions (TJJs) and the accompanying fractional Josephson effect, i.e. the presence of a $ 4\pi $ periodic energy phase relation (EPR) instead of a $ 2\pi $ periodic one, have garnered wide spread attention due to them being a possible host of Majorana zero modes \cite{Kitaev2001,Kwon2004,Fu2008,Fu2009b,Badiane2011,San-Jose2012,Dominguez2012,Beenakker2013,Virtanen2013,Houzet2013,Crepin2014,Lee2014,Kane2015,Peng2016,Kuzmanovski2016,Pico-Cortes2017,Dominguez2017,Klees2017,Cayao2017,Frombach2018,Sticlet2018,Soori2020,Hedge2020,Bernard2021,Svetogorov2021}. One of the most widely considered physical platform of such type of junctions consists of s-wave superconductors in proximity to helical edge states of two dimensional topological insulators \cite{Fu2009b,Badiane2011,Beenakker2013,Crepin2014,Bocquillon2016,Deacon2017,Frombach2018,Bernard2021}.

Majorana zero modes are quasiparticles that are their own antiparticles \cite{Kitaev2001,Read2000,Hasan2010,Qi2011,Alicea2012,Leijnse2012,Beenakker2013b,Aguado2017,Schuray2020,Prada2020}, i.e. the creation and the annihilation operator for such excitations are identical. They obey non-abelian braiding statistics and are considered as possible building blocks for topological qubits \cite{Ivanov2001,Bravyi2006,Nayak2008,Leijnse2012,Sarma2015,Lahtinen2017,Beenakker2020}. In TJJs, they appear in pairs bound to the junction and are protected at a phase difference of $\pi$ by parity conservation. The Josephson effect involving the Majorana zero modes proceeds in transport of single electrons rather than Cooper pairs which results in a $4\pi$-periodic current-phase relation despite the spectrum being $2\pi$-periodic.

Observation of the doubled periodicity is complicated by the presence of an effect called quasiparticle poisoning \cite{Fu2009b,Rainis,Leijnse2012,Lee2014}, by which the fermion parity conservation of the junction --- preserving the $ 4\pi $ periodicity --- becomes broken. In these cases, the EPR generically reduces to a $ 2\pi $ periodic relation. However, even including quasiparticle poisoning, signatures of the topological nature of the junction can be seen in the critical current of the junction in the long junction limit \cite{Beenakker2013, Crepin2014, Frombach2020} where the length of the junction $ L $ is much larger than the superconducting coherence length $ \xi $.

Typical TJJs feature lengths between the two superconductors on the order of $ L \sim 100 - 500 $ nm while superconducting gaps of the host superconductor are usually $ \Delta \sim 1 $ meV \cite{Hart2014,Knez12,Oostinga08,Wiedenmann2016}. With typical Fermi velocities $ v_F \sim 10^5 $ m/s this results in superconducting coherence lengths $ \xi = \hbar v_F / \Delta \sim 65 $ nm so that the above mentioned junctions should all lie in the long junction regime. However, since the superconducting effect is induced into the system hosting the topological edge states via the proximity effect, the induced superconducting gap is generally smaller than the gap of the host superconductor. With an induced gap that is one order of magnitude smaller than the gap of its host superconductor, $\xi$  increases to $ \xi \sim 650 $ nm so that some of the above mentioned junctions belong to the short junction regime render measurements like the one proposed in Ref.~\citenum{Beenakker2013} impossible. Another practical issue for fractional long junctions might be the enhanced susceptibility to imperfections like impurities and inelastic scattering. 

To address these shortcomings, we propose schemes in Josephson junctions based on s-wave superconductors and quantum spin Hall insulator (QSHI) edge states which are predicted to appear in graphene-like structures with intrinsic spin-orbit coupling \cite{Kane2005,Liu11}, in HgTe-based quantum wells \cite{Bernevig2006, Koenig2007, Roth2009} and their bilayers \cite{Michetti2012, Michetti2013}, in inverted type type-II semiconductors \cite{Liu2008,Knez2011,Knez2012}, in bismuthene on a SiC substrate \cite{Reis2017}, in two-dimensional TDMs \cite{Qian2014}, and on hinges of higher-order topological insulators \cite{Schindler2018, Langbehn2017,Bernard2021}. The idea we propose is to effectively tune the junction length by electrical gates. This effective length is determined by the EPR following a pattern predicted for a junction which is longer than the actual junction length $ L $. Such an effect can be achieved by coupling the helical edge states mediating the junction to additional spin degenerate nondispersive channels or to a single quantum dot. By tuning the chemical potential in the first setup or the dot energy in the second setup the effective length of the junction can be changed.

We start by setting up a general relation to determine the EPR of Andreev bound states (ABS) and define a measure for the effective length of the junction in Sec.~\ref{sec:ABS}.
We then propose two models by which the helical edge states mediating the junction are coupled to an additional channel (Sec.~\ref{sec:nonDispersiveChannel}) or a single quantum dot (Sec.~\ref{sec:quantumDot}) and discuss the resulting effective junction lengths. Subsequently, we calculate the total energy and resulting currents of the junctions in Sec.~\ref{sec:current}.
Finally, we summarize our results in Sec.~\ref{sec:conclusion}.

\section{EPR of Andreev Bound States in QSHI-based fractional JJs}
\label{sec:ABS}

At the interface between a normal conducting (N) QSHI helical edge channel and a superconducting region (S) incident electrons from the N side cannot propagate into S if their energy is below the superconducting energy gap $ \Delta $ of S due to a lack of quasiparticle states at these energies. Instead the electrons are reflected as a hole in a process called Andreev reflection \cite{Nazarov09}. During this reflection the wave function of the hole acquires a phase of $ - \arccos(E/\Delta) - \phi_i $ with $ E $ being the energy of the incident electron and $ \phi_i $ the phase of the superconductor. In the case of a SNS Josephson junction these reflected holes will again be reflected at the second superconductor as an electron. If the electron and hole pick up a phase equal to an integer multiple of $ 2\pi $ \cite{Kulik69,Houten91}

\begin{equation}
\label{eq:ABS:generalRelation}
\varphi -2 \arccos \left( \frac{E}{\Delta} \right) \pm \phi = 2\pi m,
\qquad
m \in \mathbb{Z}
\end{equation}

in an entire round trip an ABS forms. Here, $ \phi $ denotes the phase difference across the junction, $ \varphi $ the phase picked up by the electron and hole during propagation inside the normal region N and $ m $ an integer. Solving this relation for the energy $ E $ yields the EPR of the ABS. The phase $ \varphi $ picked up by the electron and hole during the propagation in the normal region is in general dependent on the ABS energy $ E $ and substantially influences the EPR. In the short junction limit, i.e. where the length between the two superconductors $ L $ is much shorter than the superconducting coherence length $ \xi = \hbar v_F/\Delta $ with the Fermi velocity $ v_F $ in the normal region N, we can approximate $ \varphi \approx 0 $ so that Eq.~\eqref{eq:ABS:generalRelation} can be solved analytically to yield the well known EPR  \cite{Nazarov09}

\begin{equation}
E = \pm \Delta \cos \left( \frac{\phi}{2} \right)
\end{equation}

in the short junction limit. In the long junction limit, ($ L \gg \xi $), the phase $ \varphi $ can not be neglected, as it scales linearly with the junction length $ \varphi = 2EL/(\hbar v_F) $. For small energies ($ E \ll \Delta $) Eq.~\eqref{eq:ABS:generalRelation} now yields a linear EPR

\begin{equation}
E = E_T \left( \pi m + \frac{\pi}{2} \mp \frac{\phi}{2} \right),
\end{equation}

with the Thouless energy $ E_T = \hbar v_F/L $. As we will show below including either nondispersive channels or a single quantum dot into which electrons can scatter the phase $ \varphi $ and therefore the resulting EPR will be altered in a way mimicking an increase in the junction length $L$. Because only the EPR is changed in a way to look like it resulted from an increased $L$, despite $ L $ being constant, we will need another way to define the effective junction length without the use of $ L $. Comparing the slope at zero energy $ \Delta/2 $ of the EPR in the short junction limit to the slope of the long junction limit $ E_T $ we note that in the long junction limit the slope will be much smaller due to $ E_T \ll \Delta $. We therefore define the JJ to effectively be in a short junction limit, if the slope at zero energy is comparable to $ \Delta/2 $ and we define the junction to be in the long junction limit, if the slope at zero energy is much smaller than $ \Delta/2 $. To this end, we differentiate Eq.~\eqref{eq:ABS:generalRelation}

\begin{equation}
\frac{\partial E}{\partial \phi}
= \mp \left[ 
\frac{\partial \varphi}{\partial E}
+ \frac{2}{\Delta \sqrt{1 - \left( \frac{E}{\Delta} \right)^2}} \right]^{-1} .
\end{equation}

\section{Modified Edge Channel}
\label{sec:nonDispersiveChannel}

\subsection{Continuum Model}

\begin{figure}
\includegraphics[width = \columnwidth]{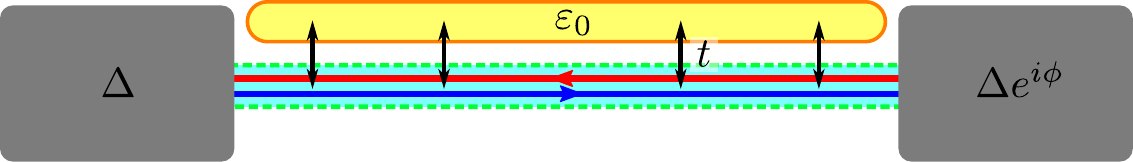}
\includegraphics[width = 0.5\columnwidth]{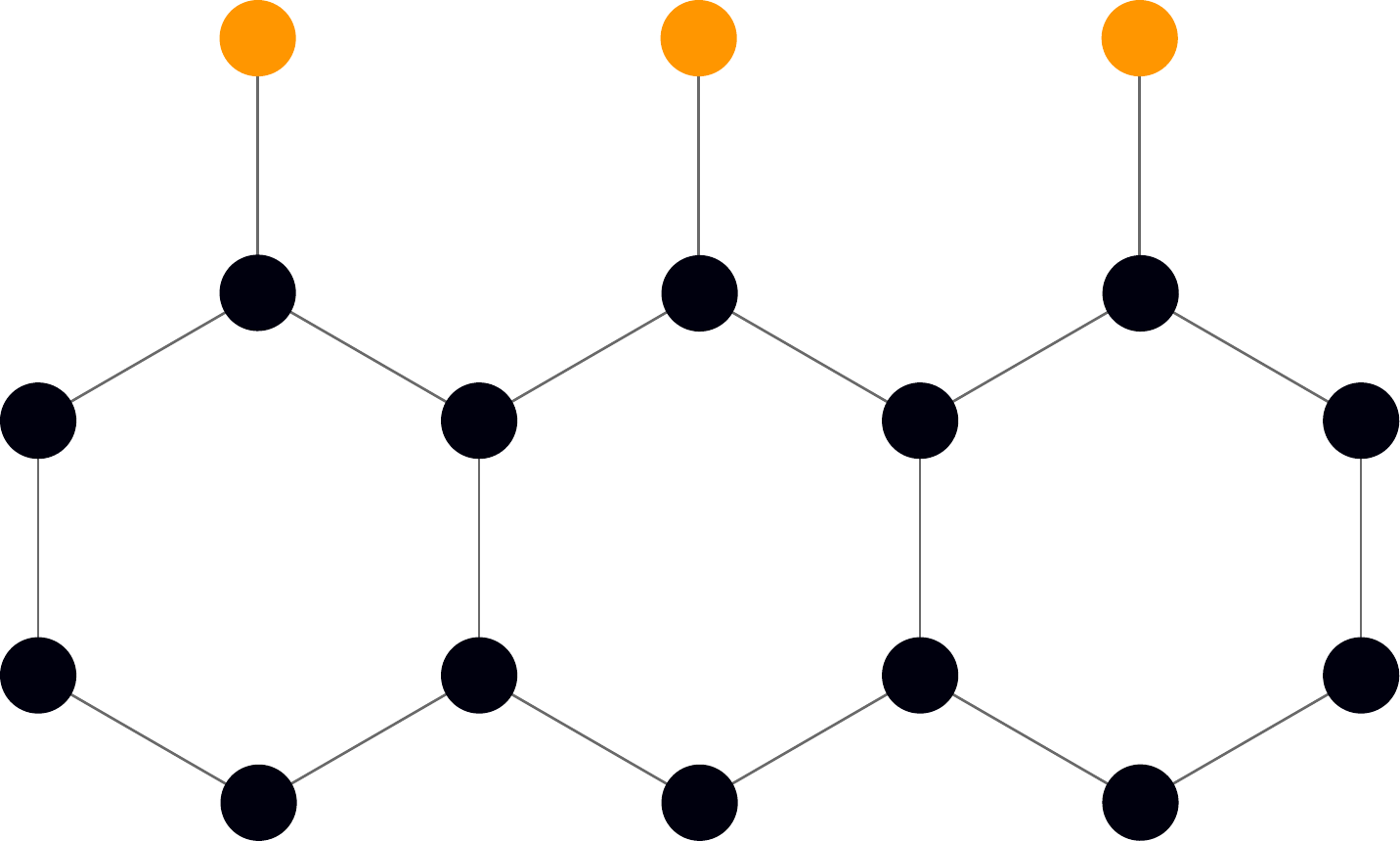}
\caption{Helical edge states (blue and red) couple to the nondispersive channels (yellow) lying at energy $ \epsilon_0 $. The two superconductors (gray) on the left and right form a Josephson junction with phase difference $ \phi $. A possible experimental realization (bottom) is offered by the Kane-Mele model in the QSHI regime, where additional atoms (yellow) are coupled to a zigzag edge\cite{Malki2017}.}
\label{fig:setupChain}
\end{figure}

A first example of a system featuring a tunable effective junction length is given by helical edge states coupled to a nondispersive spin degenerate channel (Fig.~\ref{fig:setupChain}). We consider the Hamiltonian

\begin{multline}
H = \sum_{k \sigma}
\left(\hbar v_F \sigma k - \mu\right) \psi^\dagger_{k \sigma} \psi_{k \sigma} \\
+ \left( \epsilon_0 - \mu\right) \sum_{k \sigma} d^\dagger_{k \sigma} d_{k \sigma}
+ t \sum_{k \sigma} \left(\psi^\dagger_{k \sigma} d_{k \sigma} +{\rm H.c.}\right),
\end{multline}

where $ \psi_{k \sigma} $ ($ d_{k \sigma} $) annihilates an electron with momentum $ k $ and spin polarization $ \sigma $ in the helical edge states (the nondispersive channel). The chemical potential $ \mu $ is assumed to be gate tunable while the nondispersive channel energy $ \epsilon_0 $ and the tunneling strength $ t $ (which we choose to be real for definiteness) are constant. 

Due to the momentum and spin being conserved we can write $H$ as a $2\times2$ matrix in the momentum and spin eigenbasis
\begin{equation}
H=\sum_{k\sigma}\Psi_{k\sigma}^{\dagger}\left(\begin{array}{cc} \hbar v_F \sigma k - \mu & t\\t & \epsilon_0 - \mu\end{array}\right)\Psi_{k\sigma},
\end{equation}
where $\Psi_{k\sigma}=(\psi_{k\sigma},d_{k\sigma})^{T}$.
The energy bands are then obtained by a simple matrix diagonalization. Inverting this relation yields the momentum energy relation

\begin{equation}
k = \frac{1}{\hbar v_F \sigma} \left( (E + \mu) + \frac{t^2}{\epsilon_0 - (E + \mu)} \right) .
\end{equation}

Because the phase picked up by an electron is given by $ k \cdot x $ and because the hole propagates at energy $ -E $ in the opposite direction of the electron the total phase picked up in a round trip is given by\footnote{The sign due to $ \sigma $ can be neglected, because different signs in $ \varphi $ can be absorbed in $ m $ in Eq.~\eqref{eq:ABS:generalRelation} and do not amount to new solutions of the EPR of the ABSs.}

\begin{equation}
\varphi = \frac{L}{\hbar v_F}
\left( 
2 E 
+ \frac{t^2}{(\epsilon_0 - \mu) - E}
- \frac{t^2}{(\epsilon_0 - \mu) + E}
\right),
\end{equation}

where $ E $ is the energy of the ABS. The first term is equivalent to helical edge states without scattering while the second (third) term result from electrons (holes) scattering into the nondispersive channel. The resulting slope

\begin{multline}
\label{eq:infiniteChain:slope}
\frac{\partial E}{\partial \phi}
= \pm \left[
\vphantom{\frac{2}{\Delta \sqrt{1 - \left( \frac{E}{\Delta} \right)^2}}}
\frac{L}{\hbar v_F} 
\left( 
2
+ \frac{t^2}{[(\epsilon_0 - \mu) - E]^2} 
\right. \right. \\ \left.
+ \frac{t^2}{[(\epsilon_0 - \mu) + E]^2}
\right)
\left. + \frac{2}{\Delta \sqrt{1 - \left( \frac{E}{\Delta} \right)^2}} \right]^{-1}
\end{multline}

vanishes at $ E \rightarrow \pm \Delta $ and $ E \rightarrow \pm (\epsilon_0 - \mu) $. Away from the dispersive channel resonances the slope of the case without scattering

\begin{equation}
\label{eq:without}
\frac{\partial E}{\partial \phi}
\xrightarrow{\abs{\epsilon_0-\mu\pm E} \gg t}
\pm \left[ 
\frac{2L}{\hbar v_F}
+ \frac{2}{\Delta \sqrt{1 - \left( \frac{E}{\Delta} \right)^2}} \right]^{-1}
\end{equation}

is recovered. The slope at zero energy

\begin{equation}
\label{eq:infiniteChain:SlopeZeroEnergy}
\left.\frac{\partial E}{\partial \phi}\right|_{E = 0}
= \pm \frac{\Delta}{2} 
\frac{1}{
\left( 1
+ \frac{L}{\xi} \right)
+ \frac{L}{\xi} \left[ \frac{t}{\epsilon_0 - \mu} \right]^2
}
\end{equation}

vanishes $ \sim(\mu - \epsilon_0)^2 $ for chemical potentials close to $ \epsilon_0 $ ($ \abs{\mu - \epsilon_0} \ll t $). Starting with a junction where $ L \ll \xi $ we can therefore change the junction from behaving like a short junction by tuning $ \abs{\mu - \epsilon_0} \gg \sqrt{L/\xi} t $ to a regime where the junction behaves like a long junction at low energies by tuning $ \abs{\mu - \epsilon_0} \ll \sqrt{L/\xi} t $. An effective length of such a junction at low energies can be defined as $ L_\textrm{eff} = L(1 + t^2/(\epsilon_0 - \mu)^2) $ which can be tuned to arbitrarily long lengths.

\begin{figure}
\includegraphics[width = \columnwidth]{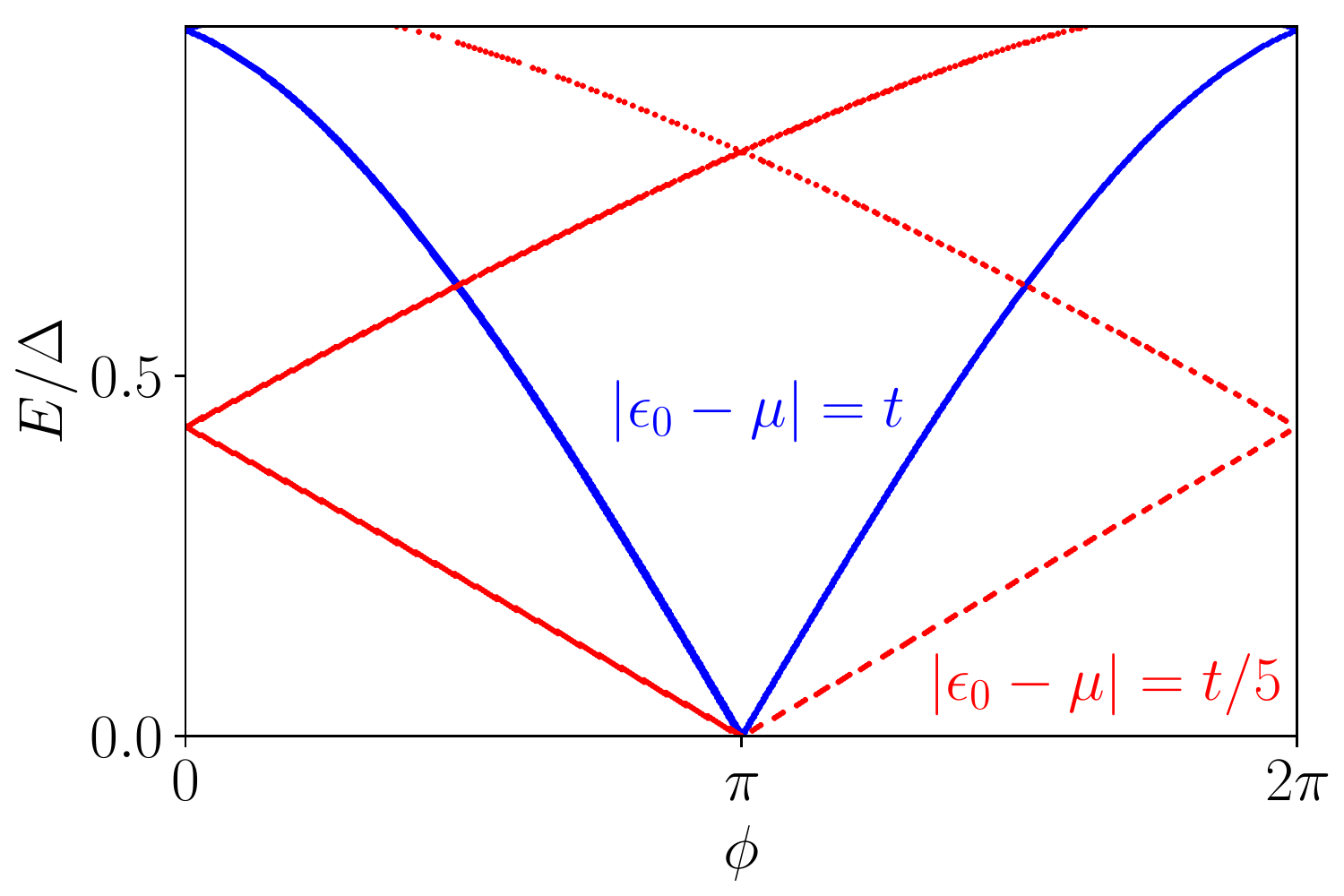}
\caption{EPR of the ABSs resulting from numerically solving Eq.~\eqref{eq:ABS:generalRelation} for a junction length $ L = 0.1 \xi $ and coupling strength $ t = 10\Delta $ for chemical potentials far away (blue) and close to (red) the channel energy $ \epsilon_0 $. The slopes around zero energy follow the relation Eq.~\eqref{eq:infiniteChain:SlopeZeroEnergy}.}
\label{fig:infiniteChainEPR}
\end{figure}

This effect can qualitatively be seen in the EPR\footnote{For low energies one can approximate $ \arccos(E/\Delta) \approx \pi/2 $. Eq.~\eqref{eq:ABS:generalRelation} then becomes an algebraic equation to third order in the energy $ E $. In general, closed solutions exist for third order algebraic equations, however, their analytic form is generally very complicated and hence not shown here.} (Fig.~\ref{fig:infiniteChainEPR}). Upon tuning the chemical potential near the nondispersive channel energy $ \epsilon_0 $ (Fig.~\ref{fig:infiniteChainEPR}, red) the slope of the EPR diminishes and the resulting EPR resembles that of a longer junction.

\subsection{Kane-Mele tight-binding Model}

\begin{figure}
\includegraphics[width = \columnwidth]{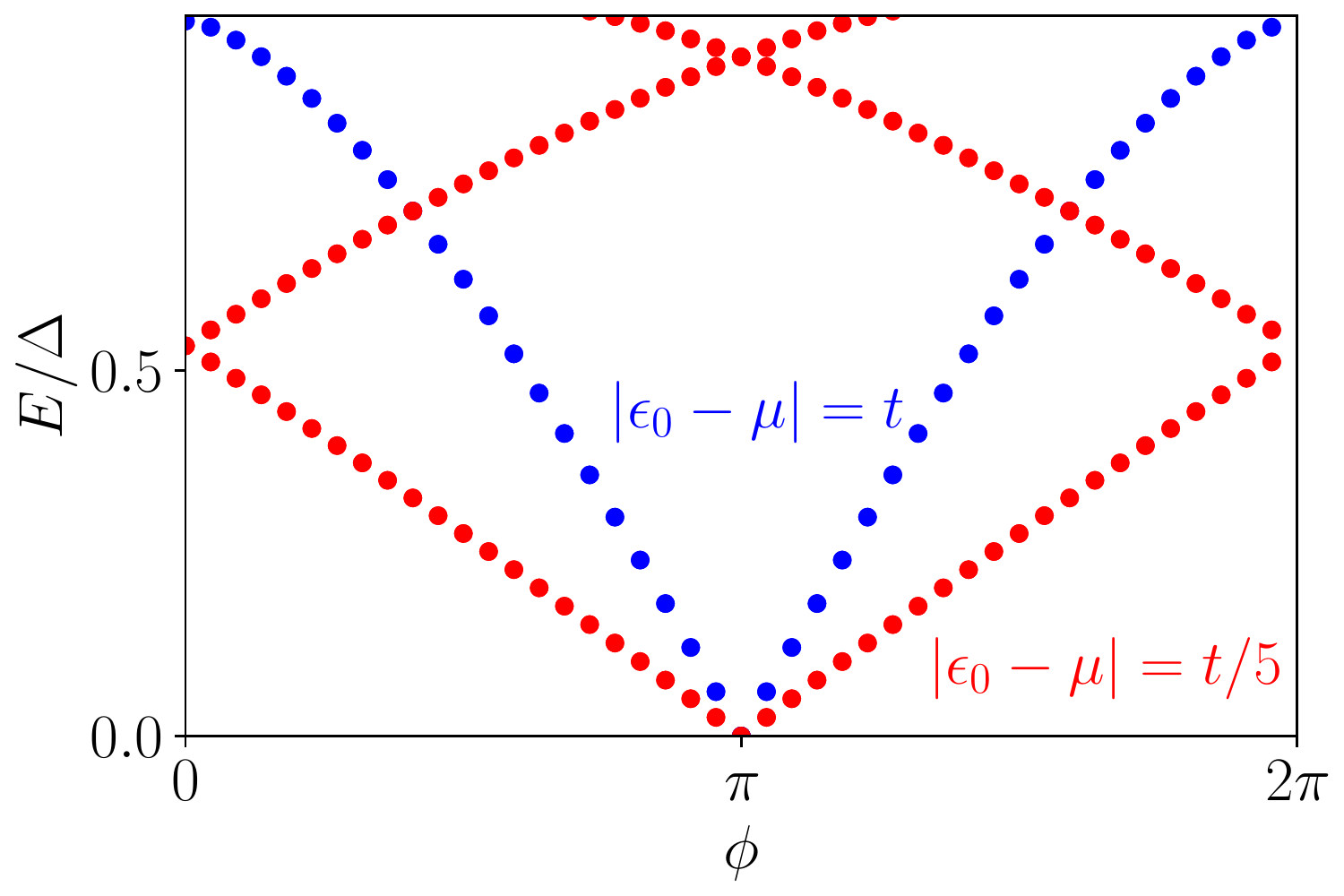}
\caption{EPR of the ABSs in the setup of Fig.~(\ref{fig:setupChain}) obtained by numerically solving a tight binding model for a junction length $ L = 0.1 \xi $ and coupling strength $ t = 10\Delta $ for chemical potentials far away (blue) and close to (red) the channel energy $ \epsilon_0 $. The slopes for the prior around zero energy is approximately $ \pm \Delta / 2 $.}
\label{fig:infinteChainEPRThightBinding}
\end{figure}

Having seen the tunability in the basic analytical model we turn our attention to a possible physical realization of such a model.
As a realization of the helical edge states we propose a Kane-Mele model in the QSHI regime which describes a single sheet of graphene, silicene, germanene or stanene \cite{Liu11}.
We consider a zigzag edge to which additional atoms are connected via a tunneling amplitude $ t $ (Fig.~\ref{fig:setupChain}, bottom). Since there exists no coupling between the extra atoms (Fig.~\ref{fig:setupChain}, bottom, yellow) these atoms form the nondispersive channel. By diagonalizing the resulting matrix the EPR can be obtained (Fig.~\ref{fig:infinteChainEPRThightBinding}). We implement the Kane-Mele model numerically in a ribbon configuration with zigzag edges to which additional atoms are coupled like described above. Two superconducting regions on both ends of the ribbon form two Josephson junctions, as the superconducting regions cover both edges of the ribbon. Further details of the implementation are given in the Appendix and Table I.

For the effectively short junction regime where the chemical potential is far away from the channel energy (Fig.~\ref{fig:infinteChainEPRThightBinding}, blue) the slope at zero energy is approximately $ \pm \Delta / 2 $ as analytically predicted.
Pushing $\mu$ closer to $\epsilon_0$  lowers the slope at $E=0$ and the junction resembles a longer junction (Fig.~\ref{fig:infinteChainEPRThightBinding}, red).
However, in this effectively long junction regime the slope at $E=0$  is larger than in the analytical case (Fig.~\ref{fig:infiniteChainEPR}, red).
Furthermore, we predicted the slope to vanish for $ \abs{\epsilon_0 - \mu} \rightarrow 0 $. In contrast to this the slope stays finite for $ \mu = \epsilon_0 $ in the tight binding model (not shown).
This discrepancy can be explained by the fact that the continuum model is only an approximation to the Kane-Mele model and, as such, we should not expect quantitatively equal results.
The effect of increasing the effective length of the junction is however observed in both systems with qualitatively similar results.

\section{Helical Edge Channel Coupled to a Quantum Dot}
\label{sec:quantumDot}

\begin{figure}
\includegraphics[width = \columnwidth]{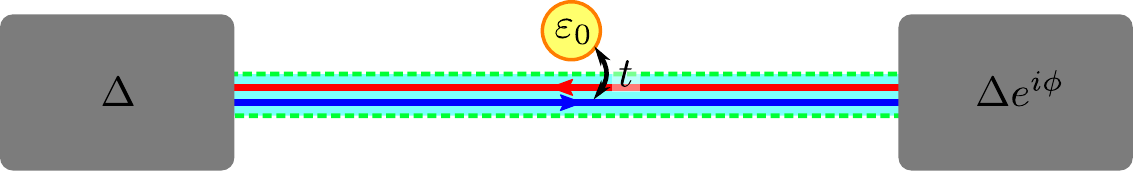}
\caption{Helical edge states (blue and red) coherently couple to the quantum dot (yellow) with spin-degenerate level energy $ \epsilon_0 $. The two superconductors (gray) on the left and right form a Josephson junction with phase difference $ \phi $.}
\label{fig:setupDot}
\end{figure}

We now move our attention to another system featuring similar gate tunable effects. We again consider a Josephson junction mediated by helical edge states (Fig.~\ref{fig:setupDot}). However, instead of coupling these helical edge states to another spin degenerate nondispersive channel like in Fig.~\ref{fig:setupChain} the helical edge states can now scatter into a quantum dot lying at the gate tunable energy $ \epsilon_0 $. The scattering is local in space and without loss of generality the dot is assumed to lie at $ x = 0 $. The system can be described with the Hamiltonian (setting $\mu=0$)

\begin{multline}
H = \sum_{k \sigma} 
\hbar v_F \sigma k \psi^\dagger_{k \sigma} \psi_{k \sigma}
+ \epsilon_0 \sum_{\sigma} d^\dagger_{\sigma} d_{\sigma} \\
+ \sqrt{L} t \sum_{\sigma} \psi^\dagger_{\sigma}(0) d_{\sigma} + {\rm H.c.},
\label{dotH}
\end{multline}

\begin{figure}
\includegraphics[width = \columnwidth]{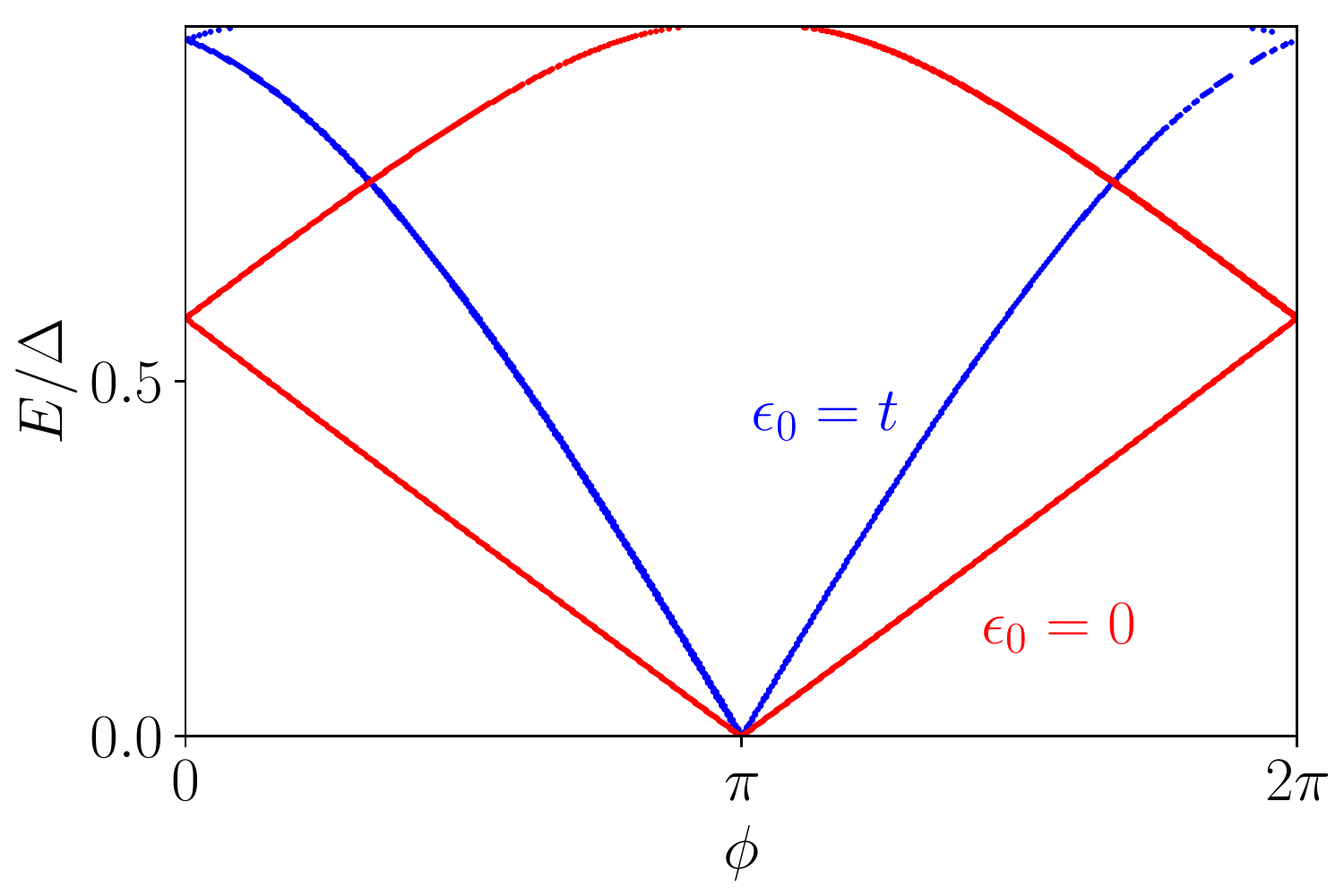}
\caption{EPR of the ABSs resulting from numerically solving Eq.~\eqref{eq:ABS:generalRelation} for a junction length $ L = 0.1 \xi $ and coupling strength $ t = 5 \Delta $ for two different dot energies. The slopes around zero energy follow the relation Eq.~\eqref{eq:singleDot:slopeZeroEnergy}.}
\label{fig:singleDotEPR}
\end{figure}

where $ \psi_\sigma(x)=(1/\sqrt{L})\sum_ke^{ikx}\psi_{k\sigma}$ and $d_\sigma$  annihilate an electron with spin polarization $\sigma$ in the helical edge states at position $x$ and on the dot, respectively. This Hamiltonian still preserves spin polarization but breaks translational symmetry. We therefore employ a scattering matrix ($S$) approach to determine the phase picked up by electrons and holes when scattering with the dot. Because $H$ in Eq.~(\ref{dotH})  preserves time reversal symmetry electrons can not backscatter and $S$ consists only of a phase factor (for forward scattering).  Utilizing the Mahaux-Weidenm\"uller formula \cite{Sumetsky19} (or the Lippmann-Schwinger equation \cite{Mello04}), the $S$-matrix element for scattering of an electron with wave number $k$ and spin $\sigma$ is expressed as
\begin{equation}
S=1+2\pi i W^{*}(H_D-E-i\pi WW^{*})^{-1}W.
\end{equation}
Here, $H_D= \epsilon_0 \sum_{\sigma} d^\dagger_{\sigma} d_{\sigma} $ and $W=\sqrt{\rho}t$ with $\rho=L/(2\pi\hbar v_F)$ the density of states describes the coupling between the channel and the dot. 
Evaluating the expression gives the phase factor $S_k = e^{i \delta}$ with

\begin{equation}
\label{eq:singleDot:scphase}
\delta = -2 \arctan \left( 
\frac{\Gamma}{E - \epsilon_0} \right),
\end{equation}
where we introduced the level width of the dot level $\Gamma=\pi\rho |t|^2$.
The phase picked up by the electron and hole in a single round trip then adds up to

\begin{equation}
\label{eq:singleDot:phase}
\varphi
= \frac{2EL}{\hbar v_F}
-2 \arctan \left( 
\frac{\Gamma}{E - \epsilon_0} \right)
-2 \arctan \left( 
\frac{\Gamma}{E + \epsilon_0} \right),
\end{equation}

where the first term is due to the propagation of the electrons and holes and the second (third) term is due to electrons (holes) scattering with the dot. The resulting slope of the EPR is given by

\begin{multline}
\frac{\partial E}{\partial \phi}
= \pm \frac{1}{2}\left[  \frac{L}{\hbar v_F}
+ \frac{ \Gamma}
{(E - \epsilon_0)^2
	+ \Gamma^2} 
\right.  \\  \left.
+ \frac{ \Gamma}
{(E + \epsilon_0)^2
	+ \Gamma^2}
 + \frac{1}{\Delta \sqrt{1 - \left( \frac{E}{\Delta} \right)^2}} \right]^{-1}
\end{multline}

which only vanishes for $ E \rightarrow \pm \Delta $. For short junctions, weak coupling to the dot or high dot energies ($ \Gamma \ll \abs{E \pm \epsilon_0} $) the second term in the denominators resulting from the scattering with the dot can be neglected and the slope of the EPR of the case where the helical edge states couple to a nondispersive channel Eq.~\eqref{eq:infiniteChain:slope} is recovered. That is, in this limit scattering due to nondispersive channels and scattering due to a single dot can not be distinguished. On the other hand for long junctions and/or strong coupling ($ L\abs{t} \gg \hbar v_F $) or high dot energies ($ \abs{E \pm \epsilon_0} \gg \abs{t} $) the fractions due to the scattering can be neglected and the slope for ABSs without scattering (Eq.~(\ref{eq:without})) is recovered. The counter intuitive result of a vanishing effect onto the EPR in the strong coupling limit can be explained by looking at Eq.~\eqref{eq:singleDot:scphase}. For strong couplings the phase picked up by a passing electron is maximized to $ \pm \pi $. However, the phase picked up by the hole will also be maximized to $ \pm \pi $ so that the overall phase picked up due to the scattering adds up to an inter multiple of $ 2\pi $ and hence has no effect on the EPR. The slope at zero energy

\begin{equation}
\label{eq:singleDot:slopeZeroEnergy}
\left. \frac{\partial E}{\partial \phi} \right|_{E = 0}
= \pm \frac{\Delta}{2}
\frac{1}{\left( 1 + \frac{L}{\xi} \right) 
+ \frac{L}{\xi} \frac{ \abs{t}^2}
{\epsilon_0^2 + \Gamma^2}}
\end{equation}

is lowered due to the scattering with the dot which again makes the junction appear to be longer than it is with an effective length $ L_\textrm{eff} = L [ 1 + \abs{t}^2/ ( \epsilon_0^2 + \Gamma^2 ) ] $. For vanishing dot energies the slope approaches a finite constant

\begin{equation}
\left. \frac{\partial E}{\partial \phi} \right|_{E = \epsilon_0 = 0}
= \pm \frac{\Delta}{2}
\frac{1}{\left( 1 + \frac{L}{\xi} \right) 
+ \frac{L}{\xi} \left( \frac{|t|}{\Gamma} \right)^2 }
\end{equation}

unlike in the case of a nondispersive channel, where the slope vanished in the limit $ \mu \rightarrow \epsilon_0 $. Because of this there exists a maximum effective length $ L_\textrm{eff}^\textrm{max} = L [ 1 + (|t|/\Gamma)^2 ] $ when tuning $ \epsilon_0 = 0 $. The effect of effectively lengthening the junction can again be seen in the EPR (Fig.~\ref{fig:singleDotEPR}).

\begin{figure}
\includegraphics[width = \columnwidth]{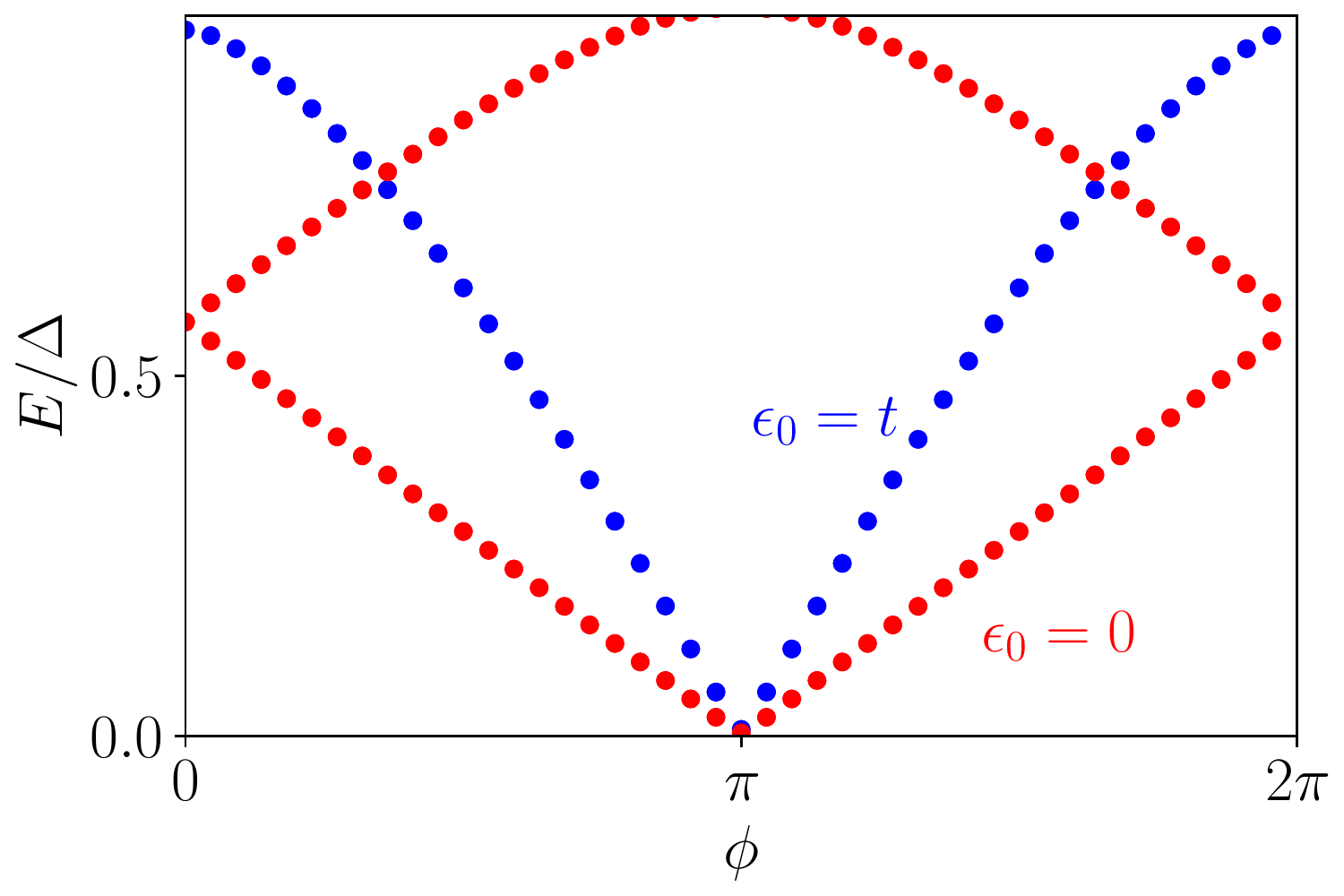}
\caption{EPR of the ABSs resulting from numerically solving a tight binding model for a junction length $ L = 0.1 \xi $ and coupling strength $ t = 5 \Delta $ for two different dot energies.}
\label{fig:singleDotEPRThightBinding}
\end{figure}

A physical realization of the quantum dot coupled to the helical edge states can be again represented by a Kane-Mele model, where only to one atom of the zigzag edge another atom is attached.
The resulting EPR (Fig.~\ref{fig:singleDotEPRThightBinding}) again follows the analytically predicted slopes.

In this section we investigated  two models analytically as well as numerically, where the helical edge channel is coherently coupled to additional channels or sites. In both situations, the coherent coupling can in practice be effectively tuned by electrical gates. We note that the coherent coupling of the helical edge channel to other quantum states (the dispersionless channel and the quantum dot) does not change the parity of the system, but merely the transmission properties of electrons and holes resulting in a modified EPR. The two Majorana zero modes are still uncoupled at $\phi=\pi$ and the zero energy states are protected by parity conservation as well as by time-reversal symmetry. The latter symmetry is present in the considered cases, however, the parity conservation would protect the Majorana zero modes even in the presence of time-reversal symmetry breaking fields \cite{Fu2009b}.

\section{Josephson Current}
\label{sec:current}

\subsection{Total Junction Energy}

In the preceding sections we focused on the effects that additional electronic states coupled to the helical edge channels have on the EPR of the ABSs.
Existing experiments however focus on derivative properties like the Josephson current.
The Josephson current at $ T = 0 $ is given by
\begin{equation}
\label{eq:current}
I = \frac{e}{\hbar} \partial_\phi E_{tot}
\end{equation}
where $ E_{tot} $ is the total energy of the system.
As we are only interested in the $ \phi $-dependent parts we can neglect the contributions from the superconducting continuum.
For the ground state this total energy is given by the sum of all single particle energy states below zero
\begin{equation}
\label{eq:totalEnergy}
 E_{tot}^{0} = \sum_{E_i \leq 0} E_i
\end{equation}
where $ E_i $ are the energies of the ABSs examined in the previous sections.
Similarly, the energy of the first excited state can be determined by adding the energy $ E^* $ of the lowest energy single particle state above 0.
However, due to the particle hole symmetry one has to simultaneously subtract the energy $ E' = -E^* $ of the highest energy state below 0 such that the total energy of the first excited state is given by
\begin{equation}
\label{eq:totalEnergyExcited}
 E_{tot}^{1} =  E_{tot}^{0} + 2E^*.
\end{equation}

\begin{figure}
\includegraphics[width = \columnwidth]{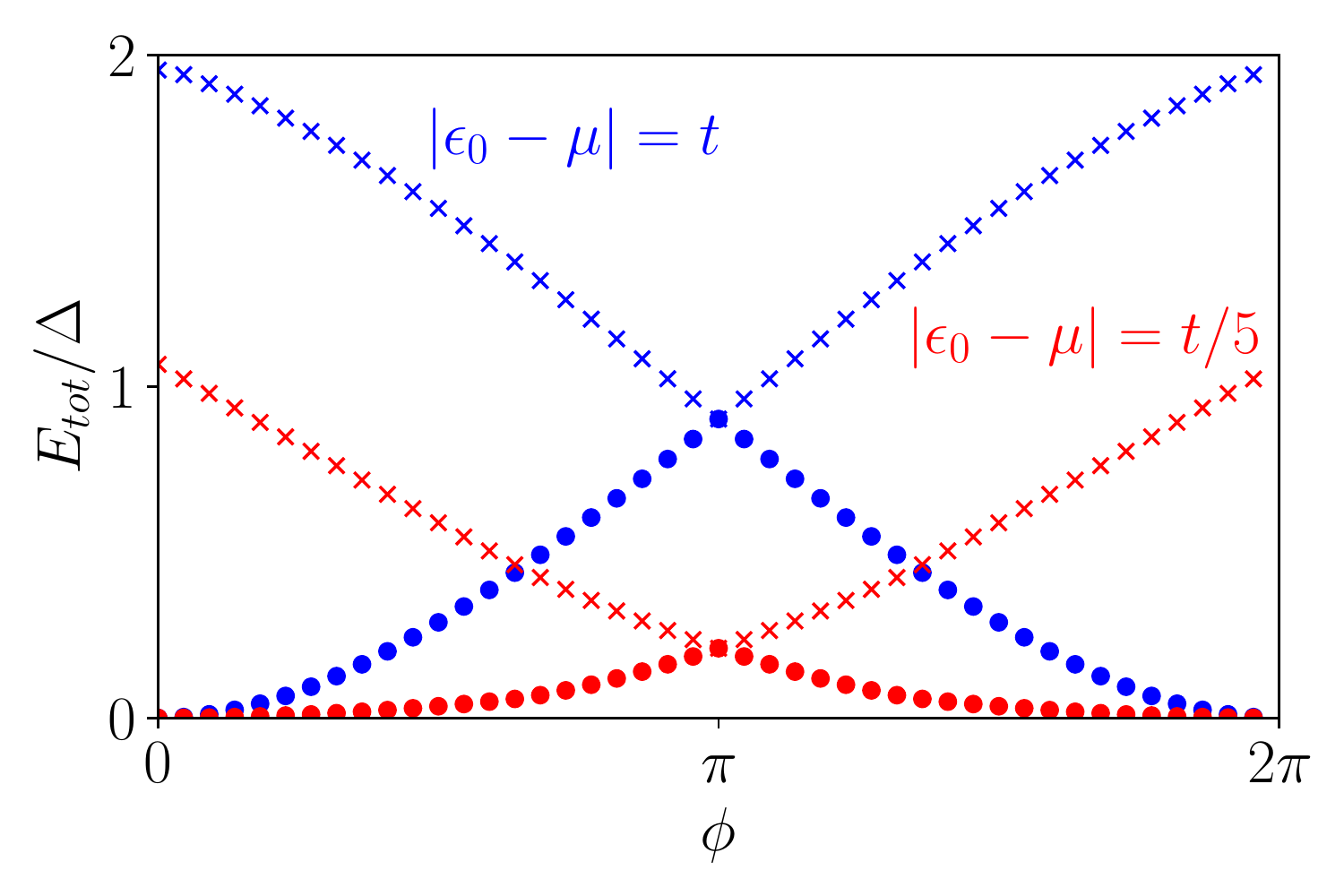}
\caption{Total energy of the system coupled to the non-dispersive channel obtained by evaluating Eq.~\eqref{eq:totalEnergy} (circle) and Eq.~\eqref{eq:totalEnergyExcited} (cross) for the single particle spectra shown in Fig.~\ref{fig:infinteChainEPRThightBinding}.
The total energy has been shifted such that $ E_{tot} = 0 $ for the ground state at $ \phi = 0 $.}
\label{fig:chainTotalEnergy}
\end{figure}

These expressions for the total energy of the ground and first excited state can be evaluated numerically for the system where an additional non-dispersive channel is coupled to the helical edge states. For the single particle spectra shown in Fig.~\ref{fig:infinteChainEPRThightBinding} the total energy undergoes a change from a cosine-like shape in the effectively short regime (Fig.~\ref{fig:chainTotalEnergy}, blue) to a quadratic shape in the effectively long regime (Fig.~\ref{fig:chainTotalEnergy}, red).
This change illustrates the junction appearing longer than it physically is in the case of chemical potentials near the channel energy previously discussed in Sec.~\ref{sec:nonDispersiveChannel}.
The energetics of the system coupled to a quantum dot (not shown) is very similar to the system coupled to the non-dispersive channel.

\subsection{Josephson Current}

\begin{figure}
\includegraphics[width = \columnwidth]{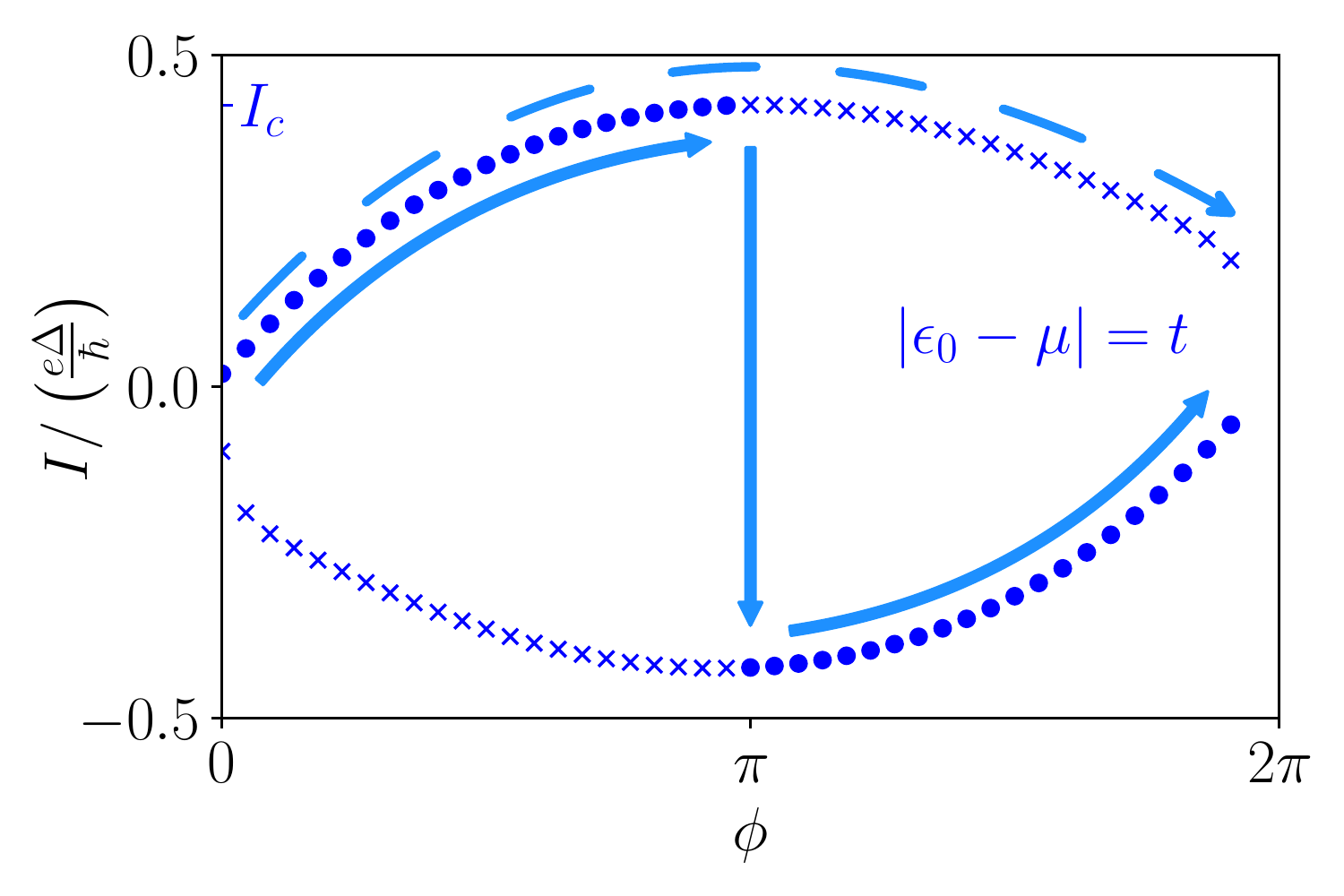}
\includegraphics[width = \columnwidth]{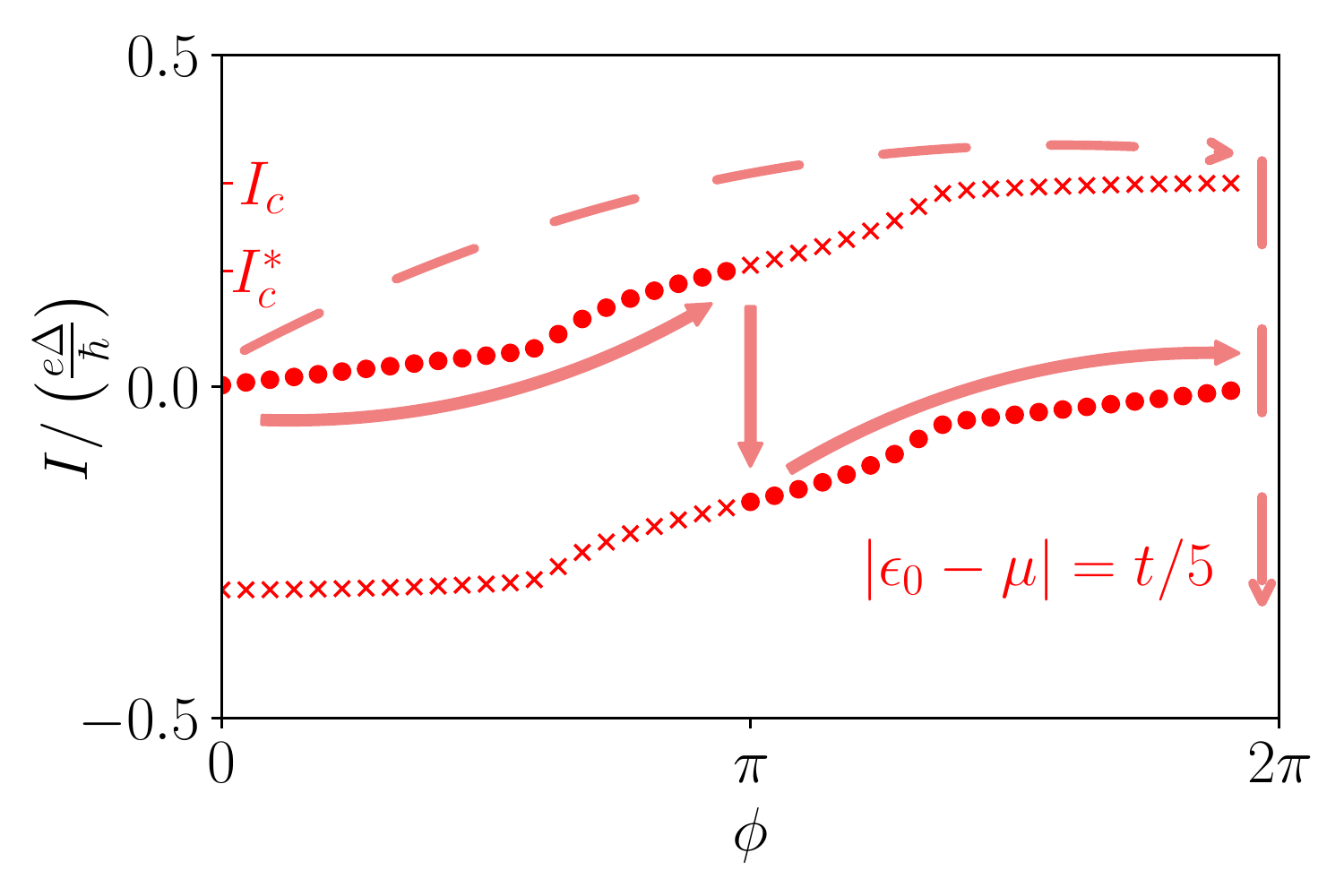}
\caption{Josephson current of the system coupled to the non-dispersive channel obtained by evaluating Eq.~\eqref{eq:current} for the ground (circle) and first excited (cross) states with a finite difference quotient for the single particle spectra shown in Fig.~(\ref{fig:infinteChainEPRThightBinding}).
The solid (dashed) arrows indicate the path the system would take in the presence (absence) of poisoning when sweeping the phase.}
\label{fig:chainCurrent}
\end{figure}

Because the total energies are only known for discrete phases, we have to approximate the derivative $ \partial_\phi E_{tot} $ with a difference quotient (forward difference) to determine the Josephson current $ I $ (Fig.~\ref{fig:chainCurrent}).
In the case where the system is coupled to an additional non-dispersive channel, tuning from the effectively short (Fig.~\ref{fig:chainCurrent}, top) to the effectively long (Fig.~\ref{fig:chainCurrent}, bottom) junction regime changes the current phase relation from a sine-shape to a more linear shape as would be expected when going from a short to a long junction.

When the parity of the junction can freely change, e.g. due to the presence of sufficiently fast quasiparticle poisoning, the system will always remain in the ground state and follow the solid arrows when sweeping the phase.
Conversely, when the crossing of the ABS spectrum at $ E = 0 $ is protected and the parity of the junction stays fixed, the junction will follow the dashed arrows when sweeping the phase \footnote{
We assume the current jumps from positive to negative values at $ \phi = 2\pi $ in the effectively long junction (Fig.~\ref{fig:chainCurrent}, bottom) as indicated by the dashed arrow on the right.
Because the ABS spectrum (Fig.~\ref{fig:infinteChainEPRThightBinding}) touches the superconducting continuum, the perfect pumping of quasi-particles should lead to a $ 2\pi $ periodic current \cite{Fu2009b}.
Although there are several ways how to observe a $4\pi$-periodic current (see e.g. Ref.~\citenum{Sticlet2018}), the value of $I_c$ is independent on such subtleties
}.
When comparing the critical currents (i.e. $I_c=|\max\limits_{\phi}I(\phi)|$) of these two cases, one can see that the critical current $ I_c $ is identical in the presence and absence of poisoning in the effectively short junction regime (Fig.~\ref{fig:chainCurrent}, top).
On the other hand, in the effectively long junction regime (Fig.~\ref{fig:chainCurrent}, bottom) a system with poisoning will result in a smaller critical current $ I_c^* $ than a system without poisoning.
Beenakker \textsl{et al.} \cite{Beenakker2013} predicted this difference of the critical current for long junctions to be exactly a factor of two.
We do not observe a factor of two difference which can be explained by the fact that this factor was only determined to be precisely equal to two in the limit of infinitely long junctions, whereas we do consider finite-size junctions.
However, the fact that only one critical current exists for short junctions regardless of the presence of poisoning while two different critical currents exist for effectively long junctions is consistent with the results of Ref.~\onlinecite{Beenakker2013}. The critical currents $I_c$ and $I_c^*$ displayed in Fig.~(\ref{fig:chainCurrent}) can be measured by current biasing the TJJ. When the bias current exceeds the critical current of the TJJ, a dc voltage develops across the junction \cite{Tinkham2004}. The onset of a dc voltage drop across the TJJ therefore relates to the size of the critical current.

\section{Conclusion}
\label{sec:conclusion}

We have proposed two models, in which the effective length of a topological Josephson junction can be tuned by external electrical gates.
In the first model, the helical edge states mediating the Josephson junction are coupled to an additional non-dispersive spin degenerate channel running along the helical edge states.
By tuning the chemical potential from far away to close to the non-dispersive channel energy the effective junction length increases with no bound.
Such a system can be realized by a Kane-Mele model with zigzag edges, where additional atoms are bound to the edge (Fig.~\ref{fig:setupChain}, bottom).
Via numerical tight binding calculations we have shown that this model effectively behaves like the analytical model, albeit with a finite maximum effective length.
In the second model, the helical edge states mediating the Josephson junction are coupled locally to a single spin degenerate quantum dot.
By tuning the quantum dot energy from far away to close to the chemical potential of the superconductors the effective junction length again increases. With this setup, a maximum possible effective junction length is reached when the quantum dot energy and chemical potential coincide.
For a possible realization of this model we again consider the Kane-Mele model with a zigzag edge, where now only one atom of the zigzag edge is coupled to one additional atom acting as the quantum dot.
Here, an increase of the effective junction length could be observed and coincided exactly with the analytical predictions.
Finally, we showed that effects from tuning the effective junction length are not limited to the ABS spectra but also have decisive signatures in the Josephson current. In particular, we argued that parity changing processes (quasiparticle poisoning) lead to a change of the critical current in effectively long junctions, whereas the critical current is insensitive to parity changing processes in effectively small junctions. The proposed effects provide a tunable Josephson junction that allow to investigate topological effects in effectively long JJs even when the physical length is comparably small.

\acknowledgments
We acknowledge useful discussions with F. Dominguez and financial support by the Deutsche Forschungsgemeinschaft (DFG, German Research Foundation) within the framework of Germany’s Excellence Strategy-EXC-2123 QuantumFrontiers-390837967.


\begin{appendix}

\section{Numerical Parameters for the Kane-Mele Model}

We implement the tight-binding versions of the helical edge states coupled to the dispersionless channel (see Fig.~1) and to a single quantum dot (see Fig.~4) by a Kane-Mele type of model \cite{Kane2005} and fix the nearest neighbor hopping energy of the honeycomb lattice to $ 2 / \sqrt{3} $. Furthermore, we set the lattice constant to one. The parameters chosen for the simulations in the main text are given in Tab.~\ref{tab:numerical_parameters}. S-wave superconductivity is implemented via the Bogoliubov-de Gennes formalism with the pairing amplitude $\Delta$ becoming an onsite-orbital contribution in the Nambu basis \cite{Frombach2018}. The ABSs are obtained numerically by matrix diagonalization of
the corresponding Bogoliubov-de Gennes Hamiltonian implemented for a SNS-junction in a ribbon geometry with a large enough ribbon width such that the contributions from the two edges become independent and degenerate. For the total energies and Josephson currents in Sec.~V, the numerical values are divided by two, which corresponds to the situation of a single edge coupled to two superconductors.

\begin{table*}
\centering	
\renewcommand{\arraystretch}{1.3}
\setlength{\tabcolsep}{7mm}	
\begin{tabular}{ @{} l l l @{} }
	\toprule
	\textbf{Parameter} & \textbf{Symbol} & \textbf{Value} \\
	\midrule
	Nearest neighbor hopping energy & & $ 2 / \sqrt{3} $ \\
	Lattice constant & & $ 1 $ \\
	\addlinespace[5mm]
	Planck constant times Fermi velocity & $ \hbar v_F $ & $ 1 $ \\
	Superconducting gap & $ \Delta $ & $ 0.01 $ \\
	Distance between superconducting leads & $ L $ & $ 10 $ \\
	System length & & $ 500 $ \\
	System width & & $ 10 $ \\
	Spin orbit interaction strength & & $ 1 $ \\
	\bottomrule
\end{tabular}
\caption{Parameters chosen for the Kane-Mele model}
\label{tab:numerical_parameters}	
\end{table*}
	
\end{appendix}
\newpage


%

\end{document}